# New assembly route for three-dimensional metamaterials through effect medium theory

Jiaguang Han [1], Sher-Yi Chiam [1], Ee Jin Teo [1], Andrew A. Bettiol [1] and Weili Zhang [2]

[1] Department of Physics, National University of Singapore, Singapore 11754, Singapore

[2] School of Electrical and Computer Engineering, Oklahoma State University, Stillwater, Oklahoma 74078, USA

E-mail: phyhanj@nus.edu.sg

**Abstract**

In this study, we illustrate the effective medium theories in the designs of three-dimensional composite metametails of both negative permittivity and permeability. The proposed metamaterials consist of coated spheres embedded in a dielectric host. Simple design rules and formulas following the effective medium models are numerically and analytically presented. We demonstrate that the revised Maxwell-Garnett effective medium theory enables us to design three-dimensional composite metamaterials through assembly of coated small spheres. The proposed approach allows for precise control of the permittivity and permeability and guides a facile, flexible and versatile way for the fabrication of composite metamaterials.

**1 introduction**

Recently, metamaterials has attracted a great deal of attention due to much experimental and theoretical interest on one hand, as well as its large potential applications on the other hand [1-9]. In metamaterials, two important parameters, effective permittivity $\varepsilon_{\text{eff}}$ and permeability $\mu_{\text{eff}}$ that determine their response to electromagnetic radiation, are simultaneously negative in a given frequency interval. Thus, they are also called "left-handed materials" or "negative index materials". The composite metamaterials are usually fabricated through embedding the periodic scattering elements in a homogeneous dielectric medium to provide an effective permittivity $\varepsilon_{\text{eff}}$ and permeability $\mu_{\text{eff}}$. One typical example is the split-ring resonator (SRR)/wire composite structure, in which SRRs and wires were constructed on either side of circuit board substrate [10,11]. The SRR/wire structures have proven to be useful in demonstrating the underlying wave propagation behavior of negative index materials. Similar designs have also been experimentally and theoretically utilized and discussed to achieve negative permittivity and negative permeability [11].

The initial planar metamaterial designs may have some limitations and the estimation of effective parameters is also complicated, and even sometime the experimental data for the transmission can not prove that both $\varepsilon_{eff}$ and $\mu_{eff}$ are indeed negative. In this work, we propose a different way to fabricate the artificial three-dimensional composite negative index metamaterial structures. Three-dimensional composite metamaterials allow the propagation of electromagnetic wave in all directions with unique properties that are distinct from conventional planar metamaterial structures and thus offer more promising applications. Electromagnetic waves interact with a composite structure containing scattering elements are often well characterized by the effective medium theories (EMTs). The proposed composite structures—obtained from assembling a mixture of binary coated spheres would enable both negative permittivity and permeability, where the assembly step is inherent to EMTs and is expedient to analysis and control of $\varepsilon_{eff}$ and $\mu_{eff}$. We shall explain the principle of composite metamaterial assembly method based on the EMTs and develop a general recipe for composite metamaterial designs, and thereby provide one assembly route for composite metamaterial structures with anticipant effective permittivity and permeability.

**2 Effective medium theories**

First of all, we would briefly sketch the most commonly used EMTs as far as necessary for setting basic idea for assembly of composite structures. Each of the EMTs not only provides a better approach to characterize effective physical parameters of composite materials, but also, more importantly, draws on one way for composite material assembly. The EMTs that are considered mostly for applications include the simple effective medium theory (S-EMT) (assembly of thin parallel plates), the Maxwell-Garnett (MG) model (assembly of small spheres), the revised MG (R-MG) model (assembly of coated small spheres), the Bruggeman (BG) model (assembly of two self-consistent mediums), and the general effective medium theory (G-EMT) (assembly of complex mixed components).

The first case of considerable practice interest is that of a regular assembly of thin parallel plates, which leads to the expression of S-EMT. We concentrate on periodic layered plates systems of the form as shown in Fig. 1(a) and assume that each plate can be described by homogeneous and isotropic permittivity $\varepsilon_1$ or $\varepsilon_2$ and permeability $\mu_1$ or $\mu_2$, respectively. Suppose that the incident field has its electric vector parallel to the plates. If the layer is sufficiently thin compared to the wavelength, the field in the plate may be considered to be uniform. Since the tangential component of electric vector $\vec{E}$ is continuous across a discontinuity surface, the effective dielectric constant of the system is given by the ratio of the electric displacement $\vec{D}$ to the electric vector $\vec{E}$ as [12]:

$$\varepsilon_{eff} = \frac{\vec{D}}{\vec{E}} = \frac{t_1\varepsilon_1 + t_1\varepsilon_1}{t_1 + t_2} = f_1\varepsilon_1 + f_2\varepsilon_2 = f_1\varepsilon_1 + (1-f_1)\varepsilon_2 \;, \tag{1}$$

where $f_1 = \frac{t_1}{t_1+t_2}$, $f_2 = \frac{t_2}{t_1+t_2} = 1-f_1$ are the volume fractions for each of the two different plates. This is the simplest case for assembly of multilayered structures and it is also can be used for an assembly of parallel and similar thin cylindrical rods.

Considering a homogeneous, isotropic medium of permittivity $\varepsilon_h$ and permeability $\mu_h$ containing small homogeneous, isotropic spheres with permittivity $\varepsilon_1$, permeability $\mu_1$ and radius $a$, as shown in Fig. 1(b). The uniform static electric field distribution in medium of $\varepsilon_h$ will be distorted by the introduction of spheres, where a dipole moment is induced inside the sphere. The system of such a form containing sphere scattering can be described by an effective medium permittivity $\varepsilon_{eff}$, which can be derived through relationship between the electric displacement $\vec{D}$, the electric field $\vec{E}$ and the polarization vector $\vec{P}$ as follows [13-14]:

$$\vec{D} = \varepsilon_h \vec{E} + \vec{P} = \varepsilon_h \frac{1+2\alpha n_0/3\varepsilon_h}{1-\alpha n_0/3\varepsilon_h} \vec{E} \;, \tag{2}$$

where the polarizability of sphere $\alpha$ is defined to be:

$$\alpha = 4\pi \varepsilon_h a^3 \frac{\varepsilon_1 - \varepsilon_h}{\varepsilon_1 + 2\varepsilon_h} \;, \tag{3}$$

and $n_0$ is the sphere number per unit volume. It follows that the effective permittivity is:

$$\varepsilon_{eff} = \frac{\vec{D}}{\vec{E}} = \varepsilon_h \frac{1+2\alpha n_0/3\varepsilon_h}{1-\alpha n_0/3\varepsilon_h} = \varepsilon_h \frac{1+2f(\varepsilon_1-\varepsilon_h)/(\varepsilon_1+2\varepsilon_h)}{1-f(\varepsilon_1-\varepsilon_h)/(\varepsilon_1+2\varepsilon_h)} \;. \tag{4}$$

Here the fractional volume occupied by the spheres is $f = n_0(\frac{4\pi a^3}{3})$. If we further write it in a more symmetric form:

$$\frac{\varepsilon_{eff} - \varepsilon_h}{\varepsilon_{eff} + 2\varepsilon_h} = f \frac{\varepsilon_1 - \varepsilon_h}{\varepsilon_1 + 2\varepsilon_h} \;, \tag{5}$$

Eq. (5) leads us to the well-known MG–EMT formula. The MG model usually describes an isotropic matrix containing spherical inclusions that are isolated from each other, such as the metal particles dispersed in a surrounding host matrix. Although it accounts for the scattering effect by small particles at the Rayleigh limit, the MG model is widely used for composite material analysis. Additionally, one can apply the MG approach to two or more inclusions of dielectric constants $\varepsilon_1$ and $\varepsilon_2$ with filling factors $f_1$ and $f_2$, respectively, in a host medium $\varepsilon_h$ as [15-16]:

$$\frac{\varepsilon_{eff} - \varepsilon_h}{\varepsilon_{eff} + 2\varepsilon_h} = f_1 \frac{\varepsilon_1 - \varepsilon_h}{\varepsilon_1 + 2\varepsilon_h} + f_2 \frac{\varepsilon_2 - \varepsilon_h}{\varepsilon_2 + 2\varepsilon_h} . \tag{6}$$

This equation provides one way to solve multi-phase composites.

A more complicated case extending the results above for the small sphere is that an effective medium theory for the coated spheres, and we call this a revised MG (R-MG) model. As shown in Fig. 1(c), we consider an inner or core sphere with permittivity $\varepsilon_1$, permeability $\mu_1$ and radius $a$ coated by outer sphere with radius $b$, permittivity $\varepsilon_2$ and permeability $\mu_2$. The coated sphere is surrounded by an external medium of $\varepsilon_h$ and $\mu_h$. For such a form, the effective permittivity $\varepsilon_{eff}$ of the whole volume medium can be obtained through the similar way as mentioned above. The scattered field can be attributed to the induced dipoles of core sphere and coating mantle, and thus the effective permittivity follows a similar formula of the MG model in Eq. (4) as [13]:

$$\varepsilon_{eff} = \frac{\vec{D}}{\vec{E}} = \varepsilon_h \frac{1 + 2f\widetilde{\varepsilon}}{1 - f\widetilde{\varepsilon}}, \tag{7}$$

while $\widetilde{\varepsilon}$ is defined to be:

$$\widetilde{\varepsilon} = \frac{(\varepsilon_2 - \varepsilon_h)(\varepsilon_1 + 2\varepsilon_2) + \eta(\varepsilon_1 - \varepsilon_2)(\varepsilon_h + 2\varepsilon_2)}{(\varepsilon_2 + 2\varepsilon_h)(\varepsilon_1 + 2\varepsilon_2) + \eta(2\varepsilon_2 - 2\varepsilon_h)(\varepsilon_1 - \varepsilon_2)}, \tag{8}$$

here the fractional volume occupied by the coated spheres is: $f = n_0(\frac{4\pi b^3}{3})$, and $\eta = \frac{a^3}{b^3}$ is the fraction of the total sphere volume occupied by the inner core. Without coating, Eq. (7) degenerates into the MG model shown in Eq. (4).

It is interested to note that a coated sphere will become *invisible* if $\widetilde{\varepsilon}$ equals zero. When $\widetilde{\varepsilon}$ is zero, the polarizability $\alpha = 4\pi \varepsilon_h \widetilde{\varepsilon} b^3$ is zero and thereby the scattered field results in zero rendering the coated particle invisible. In this case, the numerator in $\widetilde{\varepsilon}$ vanishes and then deduces that:

$$(\varepsilon_2 - \varepsilon_h)(\varepsilon_1 + 2\varepsilon_2) + \eta(\varepsilon_1 - \varepsilon_2)(\varepsilon_h + 2\varepsilon_2) = 0. \quad (9)$$

Eq. (9) further leads to the relationship between the core permittivity $\varepsilon_1$ and the mantle $\varepsilon_2$ as:

$$\varepsilon_1 = \varepsilon_2 \frac{(2+\eta)\varepsilon_h + 2(\eta-1)\varepsilon_2}{(\eta-1)\varepsilon_h + (2\eta+1)\varepsilon_2}. \quad (10)$$

We only need an appropriate choice for the permittivity of core sphere, outer coating mantle and external surrounding medium and then can take $\widetilde{\varepsilon} = 0$. Notice for this special case of invisible effect, all permittivity constants are treated as real numbers neglecting absorption although the R-MG model denoted in Eq. (7) is valid for the complex permittivity situation. On the other hand, if we let the denominator of $\widetilde{\varepsilon}$ equal zero, it would lead to the well known condition for excitation of Fröhlich mode, which plays a significant role in the investigations on the resonance of small particles. For this case, the relationship of $\varepsilon_1$ with $\varepsilon_2$ can be expressed as:

$$\varepsilon_1 = -2\varepsilon_2 \frac{(2+\eta)\varepsilon_h + (1-\eta)\varepsilon_2}{2(1-\eta)\varepsilon_h + (2\eta+1)\varepsilon_2}. \quad (11)$$

Numerous discussions relating surface excitation effect of small particles would involve in this expression and it is essential to the study of surface modes of small particles [13].

Another situation often encountered is that it is not possible to distinguish precisely between inclusions and host matrix, where both components are treated symmetrically. When we consider such an assembly case that both media of the mixture intersperses with each other with equal footing, the BG model is often utilized [13,15,17]. In the BG model, two elements are treated equally and their properties are determined self-consistently. The effective dielectric function for a two-phase system is obtained by solving the BG equation,

$$f\left(\frac{\varepsilon_1 - \varepsilon_{eff}}{\varepsilon_1 + 2\varepsilon_{eff}}\right) + (1-f)\left(\frac{\varepsilon_h - \varepsilon_{eff}}{\varepsilon_h + 2\varepsilon_{eff}}\right) = 0, \quad (12)$$

where notations $\varepsilon_1$ and $\varepsilon_h$ are the permittivity of two mixed elements, and $f$ is volume fraction of medium $\varepsilon_1$. Thus the effective permittivity of the composite is given by:

$$\varepsilon_{eff} = \frac{(3f-1)\varepsilon_1 + (2-3f)\varepsilon_h \pm \sqrt{[(3f-1)\varepsilon_1 + (2-3f)\varepsilon_h]^2 + 8\varepsilon_h\varepsilon_1}}{4}. \quad (13)$$

The detailed discussions about these four traditional or classical EMT models, for instance the validity or applications, have been well described previously [13-18]. Although some of them may in fact be questioned or improved through modern advancements, the EMTs described above have been widely used to characterize many composite systems and presented a better agreement with the measured results. It is important to notice that as filling fraction $f = 0$ or $f$ = 1, the S-EMT, MG model and BG model lead to the same result. Of course, if the topology is well specified, it is possible to determine the effective dielectric constant of composite materials directly through solving the Maxwell equations numerically.

To overcome the limitations of the above mentioned classical EMTs, a significant improvement has been recently devoted to the EMTs by D. McLachlan *et al.*[19], where a general effective medium theory was proposed. In this model, the effective dielectric function of composite system has an expression as follows:

$$f \frac{\varepsilon_1^{1/t} - \varepsilon_{eff}^{1/t}}{\varepsilon_1^{1/t} + A\varepsilon_{eff}^{1/t}} + (1-f)\frac{\varepsilon_h^{1/s} - \varepsilon_{eff}^{1/s}}{\varepsilon_h^{1/s} + A\varepsilon_{eff}^{1/s}} = 0, \quad (14)$$

where $t$ and $s$ are percolation exponents, $A = (1-p_c)/p_c$, and $p_c$ is the percolation threshold. The other notations have the same meaning as those in Eq. (4). This G-EMT model combines most aspects of both the percolations and classical EMTs, and thus can be widely used not limiting to the well-defined morphologies. The G-EMT has been used to successfully fit the experimental data for a large number of binary composite media and the most detailed explanation of this model is given in Ref. [19].

Although all these commonly used EMT models are derived from a dielectric viewpoint, it is worth noting that all these models should apply not only to the permittivity ($\varepsilon$), but also to the electrical conductivity ($\sigma$), the thermal conductivity ($K$), gaseous diffusion ($D$), and magnetic permeability ($\mu$) of mixtures [19-20]. This is especially important to allow for assembly of composite metamaterials based on the EMTs. For example, following analogical approach in the R-MG model, we can obtain an expression of effective permeability $\mu_{eff}$ for coated spheres. With each permittivity replaced with its corresponding permeability, the R-MG model for the effective permeability can be written as:

$$\mu_{eff} = \mu_h \frac{1 + 2f\tilde{\mu}}{1 - f\tilde{\mu}} \quad (15)$$

with

$$\tilde{\mu} = \frac{(\mu_2 - \mu_h)(\mu_1 + 2\mu_2) + \eta(\mu_1 - \mu_2)(\mu_h + 2\mu_2)}{(\mu_2 + 2\mu_h)(\mu_1 + 2\mu_2) + \eta(2\mu_2 - 2\mu_h)(\mu_1 - \mu_2)}, \quad (16)$$

here the permeability of inner core sphere, outer coating layer, and external surrounding medium are designated by $\mu_1$, $\mu_2$ and $\mu_h$; $\mu_{eff}$ is the effective permeability of the composite. $f$ and $\eta$ have the same meaning as those in Eqs. (7) and (8).

In this work, we shall embark on the study of coated spheres based on the above mentioned R-MG model for effective permittivity and permeability denoted by Eqs. (7), (8), (15) and (16), and show how metamaterial structures with both negative permittivity and permeability can be achieved from assembly of coated spheres through an appropriate design. The composite structure containing the binary coated spheres starting with the R-MG model is essential for us to guide a unique assembly route for metalmaterials. Although in the following sections we will focus on the coated sphere assembly configuration subject to the three-dimensional artificial composite metamaterials based on the R-MG model, it is noteworthy that each of the EMT models presented above would actually let us establish one possible assembly way to design composite metamaterials.

3 Simulations

The numerical evidence to better underline the performance of the R-MG model is now available through the full-wave numerical simulations. Fig. 2 shows the electric-field distributions and electromagnetic power-flow lines of the two-dimensional full-wave simulations for three cases: (a) small sphere as designed in Fig. 1(b), (b) small sphere of the same geometry as case (a) but with coating mantle, and (c) the same geometry as case (b) but with proper design in permittivity to show "invisible" effect, respectively. In these cases, the full wave electromagnetic simulations were performed by the finite element method. A 2 GHz transverse-electric (TE) plane wave is incident along the x-direction upon the small sphere or coated sphere of a diameter less than wavelength. The computational domain was terminated by perfectly matched layers (PMLs). The color map depicts the spatial distribution of the electric field oriented along the z-direction. In Fig. 2(a), we consider a small nonmagnetic sphere of permittivity $\varepsilon_1 = 5.0$ in the air. As can be seen, electromagnetic waves are scattered upon the small sphere. Fig. 2(b) extends the situation above from a small sphere to a coated sphere, where the coating shell is of permittivity of $\varepsilon_2 = 2.0$ and the core is same as in Fig. 2(a). It is noticeable how the presence of the coating shell enhances the total scattering of the plane-wave field. In particular, Fig. 2(c) represents how the performance of the coated sphere drastically reduces the total scatting of the incident plane wave and thereby renders a homogeneous coated spherical particle invisible through a proper design according to the R-MG model. Supposing the surrounding medium $\varepsilon_h = 1.0$ and coating mantle $\varepsilon_2 = 2.0$, we have core permittivity $\varepsilon_1 = -1.69$ by Eq. (10). As shown in Fig. 2(c), the invisible effect is clear with minimum scattering as expected and it is also evident from this figure how the EMTs can give desired effective physical parameters through proper designs. Compared with recent proposed cloaking [21-23], where electromagnetic fields can be bended and stretched through specified coordinate transformation, Fig. 2(c) demonstrates another route to approach cloaking through assembly of coated spheres. Actually, the scattered wave is attributed to the induced dipoles of the coated sphere, but this does not matter whether one assigns this scattering to permittivity or to permeability. In fact, if we replace all permittivities in Fig. (2) to its corresponding permeabilities, the same simulated profiles can be obtained. For the whole volume medium containing coated spheres, the scattering accounts for its effective permittivity and permeability.

Since the effective permittivity and permeability of the composite containing coated spheres are adjustable in terms of the R-MG model, one is able to construct the composite metamaterial structure of both negative permittivity and permeability through proper design in filling fraction $f$, $\eta$, permittivity and permeability of surrounding medium, core sphere, and mantle material. For instance, with fixed filling fraction $f$ and $\eta$, the effective permittivity and permeability was calculated by the R-MG model denoted in Eqs. (7) and (15) for different ratio of the core sphere permittivity or permeability to coating's values, as well as different ratio of the coated sphere permittivity or permeability to the surrounding medium's. Fig. 3 plots the calculated results for these different configurations of various material parameters to explore the sensitivity and reliability of the metamaterial composite by assembling coated spheres. It is seen that the necessary condition of realizing negative permittivity/permeability requires that the coating shell and core sphere at lest have the opposite signs in its permittivity/permeability if one assumes that the surrounding medium has a positive magnitude.

The R-MG model provides a conceptually simple approach to the design of metamaterials and it is perceivable that one can discuss various complex metamaterial configurations. As an typical instance, let us consider a fairly interest case that a metamaterial structure is fabricated by embedding coated spheres into a transparent medium $\varepsilon_h$, where a magnetic core is coated by one metal shell. Fig. 4 shows the simulated distribution of electric field, where we assumed that the core magnetic sphere is of $\varepsilon_1 = 2.0$ and, $\mu_1 = -50.0 + i100.0$ and coating metal is of $\varepsilon_2 = -24.0 + i0.5$ and $\mu_2 = 1.0$. As shown in Fig. 4, the scattering effect is seen clearly due to the induced dipoles of magnetic core and metal shell. However, a general problem is that the permeability of magnetic particle and permittivity of the metal mantle are mostly the function of the frequency. Thus, we need to take a frequency-dependent permittivity and permeability into account.

For the outer metal coating layer, it is known that the permittivity as a function of frequency can be well described by the Drude model as:

$$\varepsilon_2 = 1 - \frac{\omega_p^2}{\omega^2 + i\gamma\omega}, \tag{17}$$

where the key parameters describing the dynamics of free electrons or plasmons in a metal are plasma frequency $\omega_p = \sqrt{Ne^2/(\varepsilon_0 m^*)}$ and the carrier damping constant $\gamma$, where $N$ is the carrier density, $m^*$ is the electron effective mass, and $\varepsilon_0$ is the free-space permittivity constant with a value $8.854 \times 10^{-12}$ F/m. On the other hand, the frequency-dependent permeability of the inner magnetic particle is given by [7]:

$$\mu_2 = 1 - \frac{\omega^2}{\omega^2 - \omega_m^2 + i\Gamma\omega}, \tag{18}$$

where $\omega_m$ is the magnetic resonance frequency and $\Gamma$ is the damping constant.

Because the size of a particle in wavelength is more fundamental for light scattering than is in absolute terms, we introduce a scaling parameter: $\varDelta = b/c$, in which $b$ is the radius of the whole coated particle and $c$ is the light speed in vacuum. All the physical parameters $\omega_p$, $\omega_m$, $\gamma$, $\Gamma$ and $\omega$ used in the following calculation are normalized by $\varDelta$. Fig. 5(c) illustrate explicitly how the effective permittivity and permeability of the composite medium changes as a function of normalized frequency, where the magnetic particle coated with a metal layer is embedded into one transparent medium $\varepsilon_h$=2.33 with fixed filling factor parameters. Figs. 5(a) and 5(b) show the corresponding frequency-dependent permittivity and permeability of the core magnetic particle and outer coating metal layer, respectively. It can be found from Fig. 5(c) that the composite structure possesses both negative permittivity and permeability when normalized frequency is between 2.34 and 2.51. The optimization might be achieved, for instance, by changing the filling fraction of coated spheres, the ration of radius between the coating particle and core, or the surrounding medium. In this configuration, the composite metamaterial with a flexible permittivity and permeability can be obtained. The figure shows a practical strategy on how to use magnetic particles coated with metal shell in a proper design to realize real three-dimensional composite metamaterials.

**4 Conclusions**

In summary, we have proposed a method to fabricate three-dimensional composite metamaterials through assembly of coated spheres based on the effective medium theories. During the process it becomes obvious that the permittivity and permeability properties of composite structures are much better controlled and obtained, and may provide comparable flexibility. Therefore, it is a promising route for realization of real three-dimensional metamerials of both negative permittivity and permeability. Apart from metamaterial design, a further understanding of all presented EMT models will immediately yield new ideas for novel structure assembly and device applications.

**Acknowledgement**

The work was partially supported by the MOE Academic Research Fund of Singapore and the Lee Kuan Yew Fund.

**Figure Captions**

**FIG. 1.** Sketch of assembly of (a) thin parallel plates, (b) small sphere and (c) coated small sphere.

**FIG. 2.** Simulated electric field distribution in the external surrounding medium ($\varepsilon_h$= 1.0) containing (a) small dielectric sphere of permittivity $\varepsilon_1$= 5.0, (b) coated sphere with permittivity of a core sphere $\varepsilon_1$= 5.0 and permittivity of outer coating shell $\varepsilon_2$= 2.0, and (c) "invisible" coated sphere with permittivity of core sphere $\varepsilon_1$= -1.69 and permittivity of outer coating shell $\varepsilon_2$= 2.0. The power flow lines (grey lines) are directed in $x$ direction.

**FIG. 3.** Effective permittivity or permeability for different configurations with various permittivity or permeability components is calculated by the R-MG model for assembly of coated spheres.

**FIG. 4.** Electric field distribution in the external surrounding medium of $\varepsilon_h$= 1.0 and $\mu_h$= 1.0 containing a coated magnetic sphere. The core magnetic sphere of $\varepsilon_1$=2.0 and $\mu_1 = -50.0 + i100.0$ and the outer metal shell of $\varepsilon_2 = -24.0 + i0.5$ and $\mu_2$= 1.0 are assumed in the calculation.

**FIG. 5.** Illustration of assembly of composite metamaterials by embedding coated spheres into a transparent dielectric medium of $\varepsilon_h$= 2.33. (a) the permittivity and permeability of the magnetic inner sphere as a function of the normalized frequency. The frequency-dependent permeability $\mu_1$ is calculated by Eq. (18) with: $\omega_m \cdot \Delta$ = 2.0 and $\Gamma \cdot \Delta$ = 0.05. The permittivity is fixed as $\varepsilon_1$= 1.5. (b) the permittivity and permeability of outer metal shell as a function of the normalized frequency. The frequency-dependent permeability $\varepsilon_2$ is calculated by Eq. (17) with: $\omega_p \cdot \Delta$ = 8.5 and $\gamma \cdot \Delta$ = 0.6. The permeability is fixed as $\mu_2$= 1.0. (c) the effective permittivity and permeability of the composite as a as a function of the normalized frequency calculated by the R-MG model. The filling fraction parameters are: $f$=0.4, $\eta$=0.3.

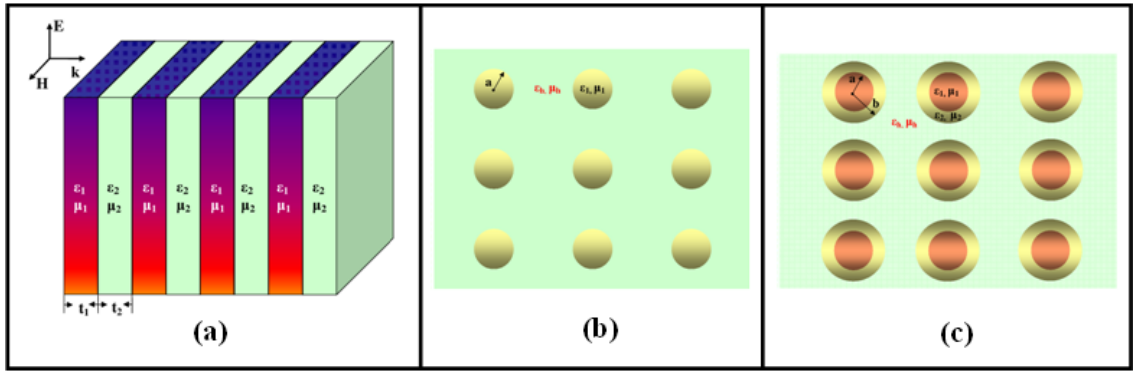

**FIG. 1.**
**Han** *et al.*

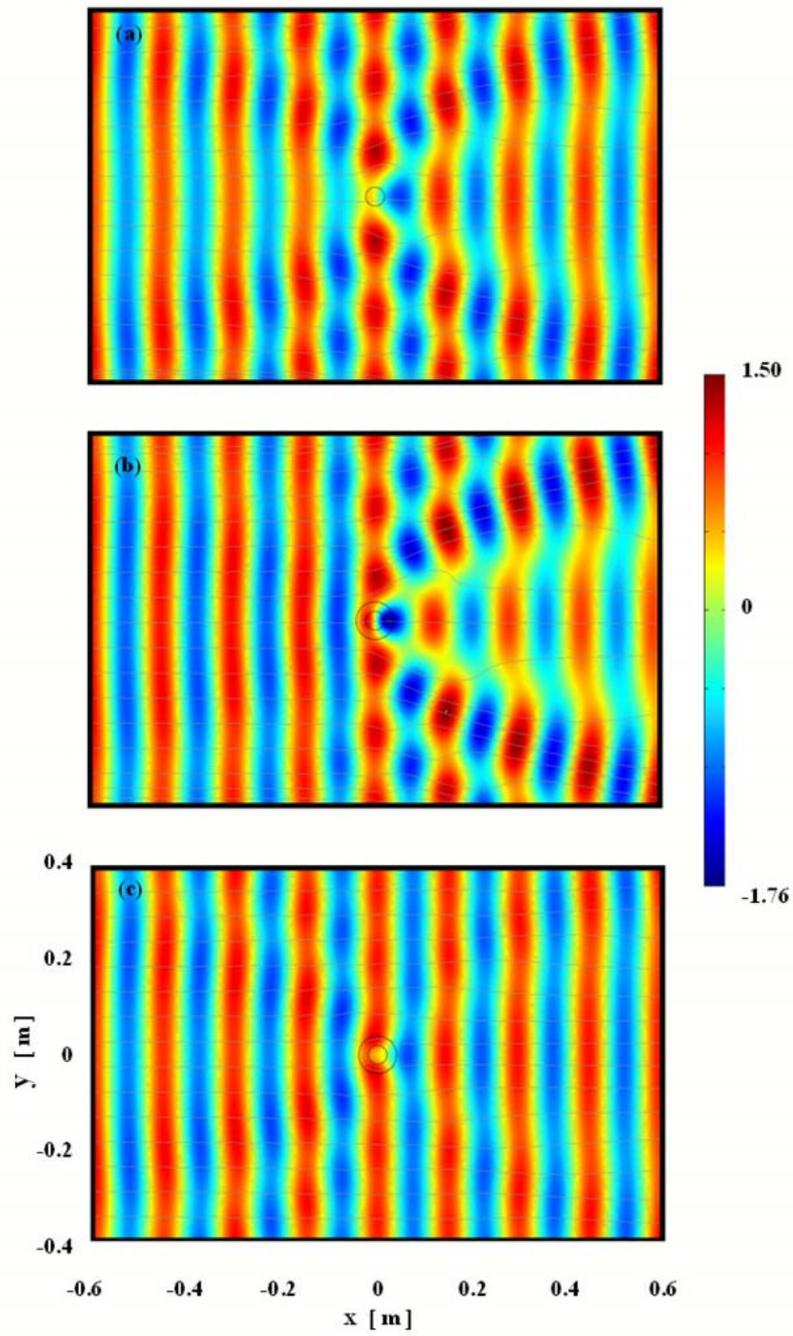

**FIG. 2.**
**Han** *et al.*

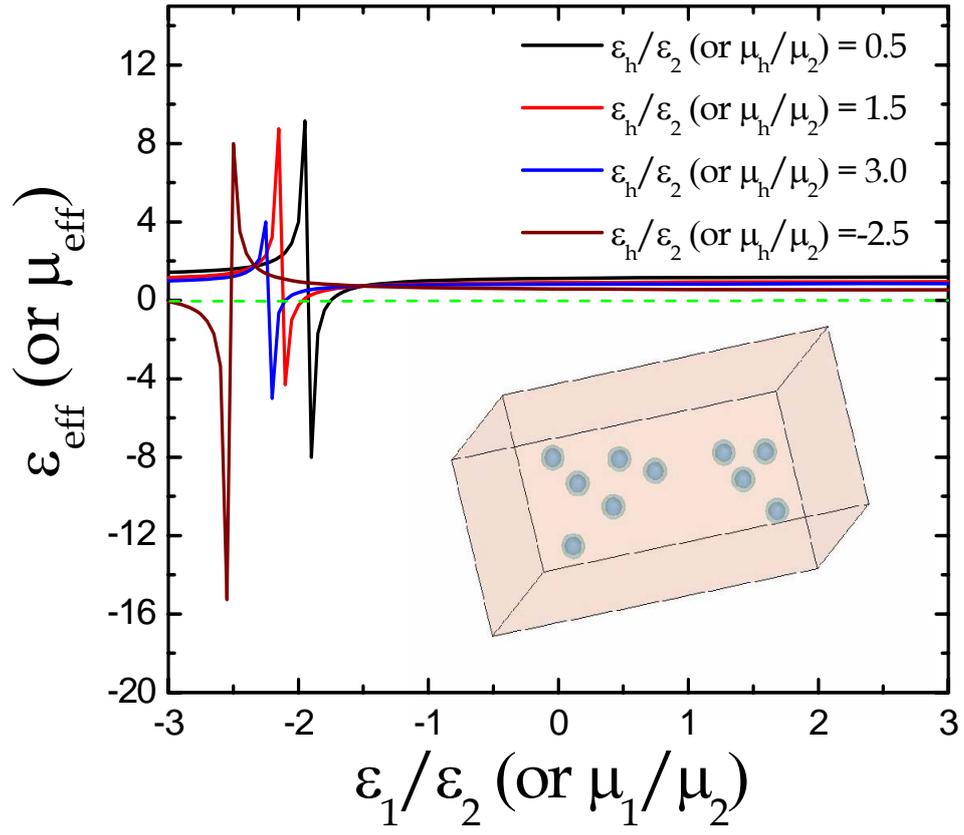

**FIG. 3.**
**Han** *et al.*

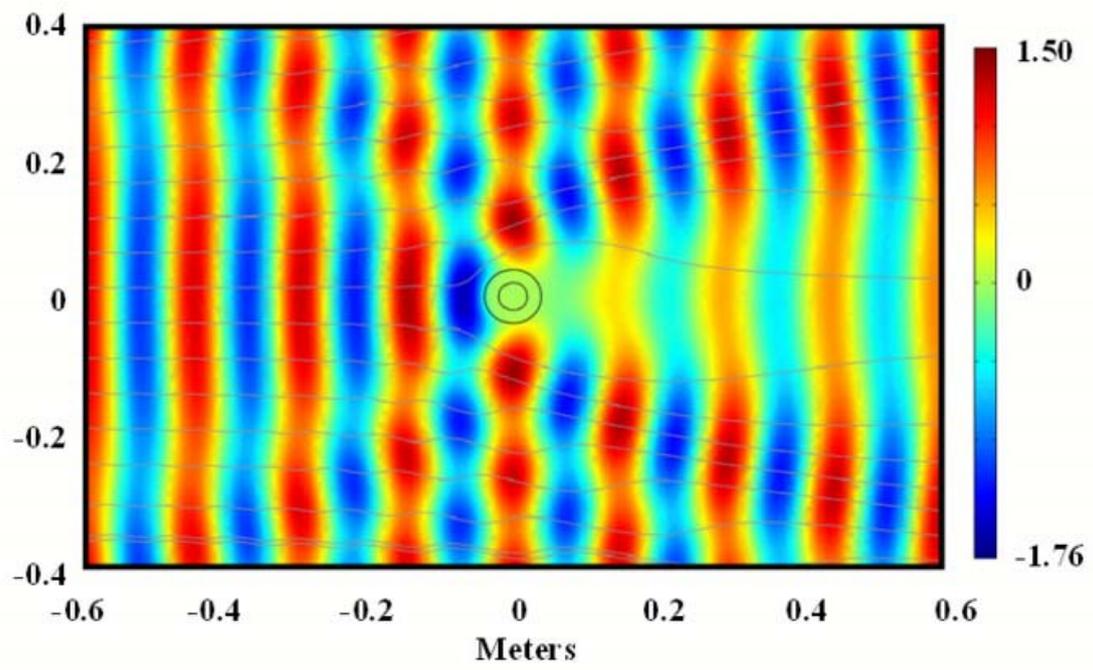

**FIG. 4.**
**Han** *et al.*

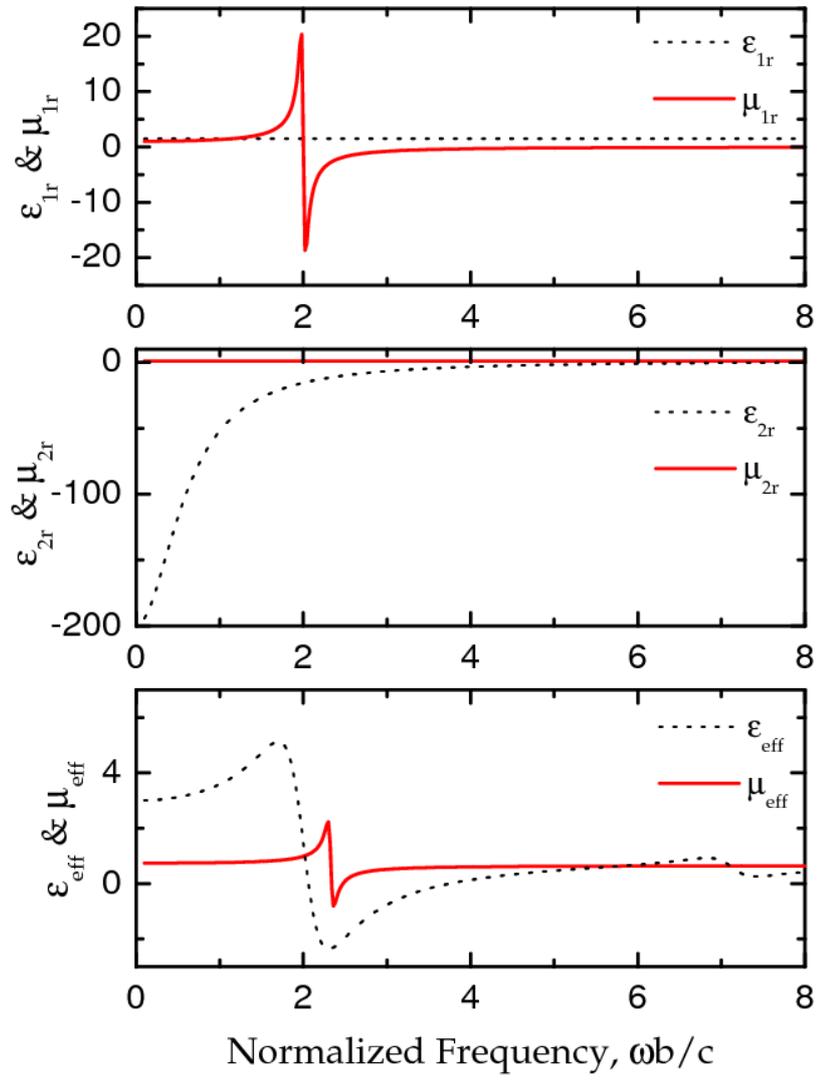

**FIG. 5.**
**Han** *et al.*